# Graph analysis of spontaneous brain network using EEG source connectivity

Aya Kabbara [a,b,c,d*]   Wassim El Falou [a,b]   Mohamad Khalil [a,b]   Fabrice Wendling [c,d]   Mahmoud Hassan [c,d]

[a] Azm research center in biotechnology, EDST, Lebanese University, Lebanon
[b] CRSI research center, Faculty of engineering, Lebanese University, Lebanon
[c] Université de Rennes 1, LTSI, F-35000, France
[d] INSERM, U1099, Rennes, F-35000, France
aya.kabbara@etudiant.univ-rennes1.fr

*Abstract*— Exploring the human brain networks during rest is a topic of great interest. Several structural and functional studies have previously been conducted to study the intrinsic brain networks. In this paper, we focus on investigating the human brain network topology using dense Electroencephalography (EEG) source connectivity approach. We applied graph theoretical methods on functional networks reconstructed from resting state data acquired using EEG in 14 healthy subjects. Our findings confirmed the existence of sets of brain regions considered as 'functional hubs'. In particular, the isthmus cingulate and the orbitofrontal regions reveal high levels of integration. Results also emphasize on the critical role of the default mode network (DMN) in enabling an efficient communication between brain regions.

## I. Introduction

Our brain is a complex network. It consists of distinct regions which are anatomically and/or functionally connected [1][2][3]. Over the past decade, several studies were interested in exploring the human brain functional organization during rest. In this context, a number of networks termed as "Resting State Networks (RSNs)" have been revealed and found to be consistent over subjects and modalities [1][4][5][6][7][3][8][9][10][11]. The networks that are frequently reported are the default mode network (DMN), the dorsal attention network (DAN), the ventral attention network (VAN), the salience network (SAN), the motor network, the visual network and the auditory network.

The high level of RSNs connectivity suggests the existence of a set of crucial regions (hubs) particularly important in providing an efficient communication between brain regions. This idea has been supported by numerous structural [12][13][14][15][16] as well as functional studies [17][18][19][20]. The most common identified regions include the cingulate region and the medial frontal region. Furthermore, [20] has demonstrated that these "hubs" are highly interconnected with each other forming a "rich-club" organization of the human brain network.

The RSNs have been explored using different neuroimaging techniques such as functional magnetic resonance imaging (fMRI), magnetoencephalography (MEG), positron emission tomography (PET) [21][22][23][20]. However, exploring the human brain architecture using electroencephalography (EEG) recordings has not been well established yet, which is the main purpose of the presented paper. For this end, we collected dense-EEG data from 14 subjects at rest with eyes closed. We then reconstructed the functional networks using the EEG source connectivity approach [24]. This step has been followed by a graph quantification of the constructed networks. Our results demonstrate the existence of crucial nodes located in the cingulate region and in the medial frontal region confirming the previous results obtained by fMRI and MEG analyses [15][16][17][18][19][20]. Moreover, the results insist on the importance of DMN in establishing efficient brain connectivity, since the majority of the identified central nodes belong particularly to the DMN.

## II. Materials and methods

### A. Data acquisition

Fourteen healthy subjects were asked to relax with their eyes closed without falling asleep, while 10 minutes of EEG were recorded. For each subject, the individual structural MRI was acquired in addition to dense EEG (256-channels, EGI, Electrical Geodesic Inc.). EEGs were sampled at 1000 Hz, band-pass filtered within 3-45 Hz. The acquisition was performed following the procedure approved by the National Ethics Committee for the Protection of Persons (CPP) (BrainGraph study, agreement number 2014-A01461- 46, promoter: Rennes University Hospital). All subjects gave an informed consent prior to their participation.

### B. Task, procedure and design

We segmented the EEGs into non-overlapping 40s epochs. A segment with amplitude ±80μV was considered as artifactual and rejected after visual inspection.

### C. Cortical network reconstruction

The reconstruction of functional networks from scalp EEGs included three main steps:

C.1) Solving the EEG inverse problem: The EEG signals S(t) are expressed as linear combinations of time-varying current dipole sources D(t): $S = G.D + B$ where G and B(t) are respectively the matrix containing the lead fields of the dipolar sources and the additive noise. The inverse method aims at estimating the parameters of the dipolar sources $\hat{D}(t)$ (notably the position, orientation and magnitude). Among the methods available, we chose to use the weighted minimum norm estimate method (wMNE) implemented in Brainstorm [25]. This method was chosen in the presented work based on a comparative study reported in [26], where authors have demonstrated the robustness of wMNE over other methods.

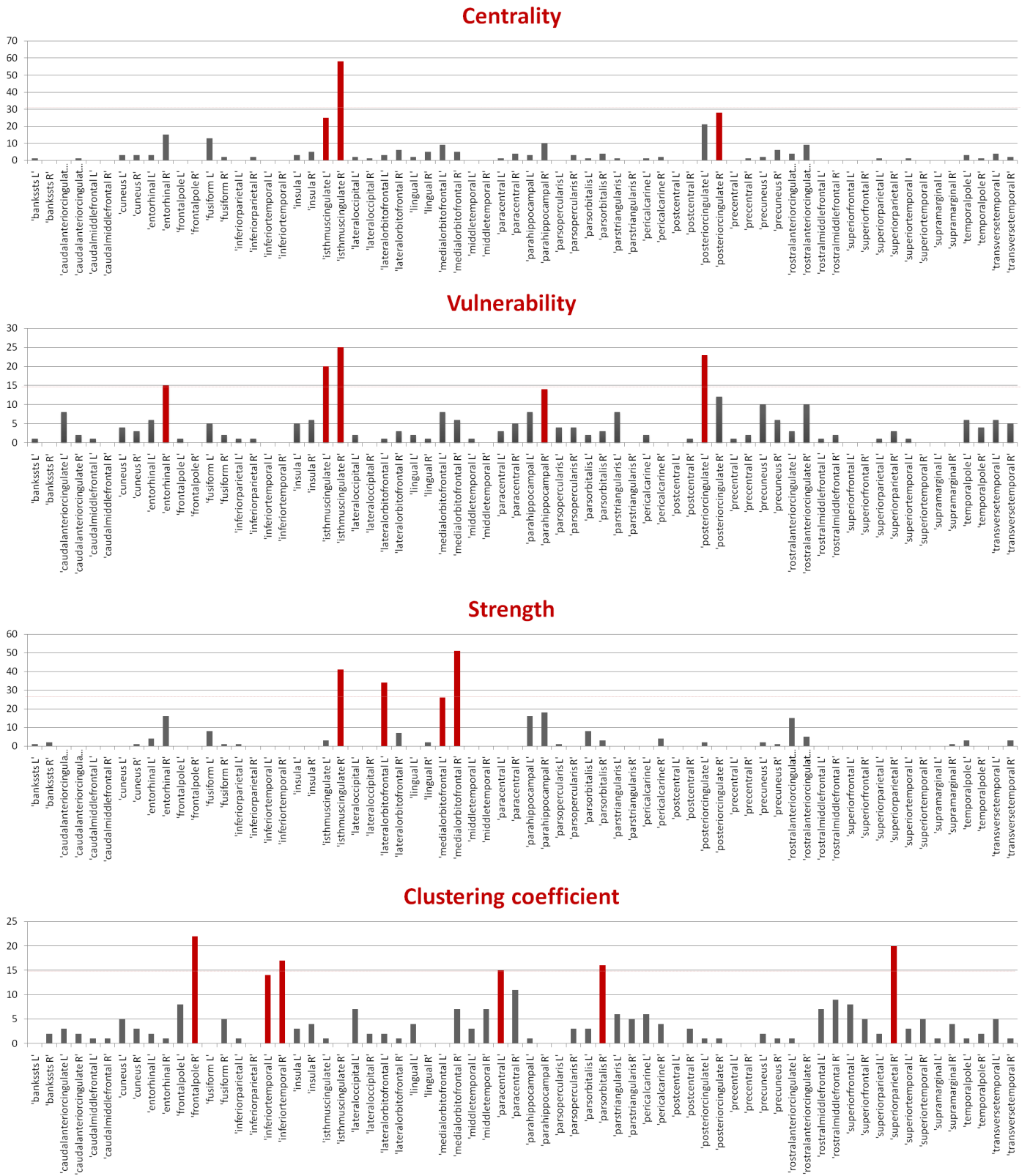

Figure 1. Distribution of the four graph measures on the 68 ROIs. The dotted line illustrated in each histogram presents the sum of the average value and twice the standard deviation of nodes distribution. A bar is colored if its value is above the dotted line.

C.2) Identifying the regions of interest (ROIs): We used the Desikan-Killiany atlas to anatomically segment the brain into 68 cortical regions [27]. Dynamics of sources according to the same ROI were averaged over time resulting in 68 regional time series.

C.3) Measuring the functional connectivity: We quantified the synchronization between the 68 regional time series using the phase locking value (PLV). PLV has been also chosen based on the comparative study performed by [26], who concluded that the wMNE/PLV combination provides the best results among many possible inverse/connectivity combinations. This combination was also recently used to track dynamics of functional brain networks during cognitive task [24]. The final networks were obtained by applying a proportional threshold (10%) to remove weak connections from the PLV matrices.

### D. Graph metrics extraction

We quantified the importance of each node in terms of four network metrics:

1- Betweenness Centrality:

$$BC_i = \sum_{i,j} \frac{\sigma(i,u,j)}{\sigma(i,j)}$$

where $\sigma(i,u,j)$ is the number of shortest paths between nodes $i$ and $j$ that pass through node $u$, $\sigma(i,j)$ is the total number of shortest paths between $i$ and $j$, and the sum is over all pairs $i,j$ of distinct nodes.

2- Vulnerability: The vulnerability of a node can be defined as the reduction in the efficiency of the network when the node and all its edges are removed:

$$V_i = \frac{E - E_i}{E}$$

Where $E$ is the global efficiency of the network before any attach, and $E_i$ is the global efficiency of the network after attacking the node $i$ [28].

3- Strength:

$$S_i = \sum_j w_{ij}$$

Where $w_{ij}$ is the weight of the edge linking the node $i$ to the node $j$.

4- Clustering coefficient: The clustering coefficient of a node in a graph quantifies how close its neighbors are to being a clique.

### III. RESULTS

For each segment of each subject, we extracted nodes with highest centrality, strength, vulnerability, and clustering coefficient values. Since the analysis was done for N=14 subjects, a node can be designated as the most critical node (in terms of centrality, vulnerability, strength and clustering coefficient) for a number of times varying from 0 to 14. The four histograms shown in Figure 1 depict the number of times each of the 68 nodes was considered as the most important node in terms of centrality, vulnerability, strength and clustering coefficient. A significant bar is colored if its value exceeds the sum of the average value and twice the standard deviation of nodes distribution. As illustrated, the most central nodes were the "left/right isthmus cingulate" and the "right posterior cingulate". The most vulnerable nodes were the "left/right isthmus cingulate", the "left posterior cingulate", the "right parahippocampal", the right "entorhinal". According to strength metric, we observe that the "right isthmus cingulate", "left/right medial orbitofrontal", the "left lateral orbitofrontal" are the most important nodes. While the right "frontal pole", the "left/right inferior temporal", the "left paracentral", the "right parsorbitalis" and the "right superior parietal" were significant with regard to the clustering coefficient.

The spatial distribution of node centrality, vulnerability, strength and clustering coefficient on the cortical surface across all participants is shown in figure 2. The figure shows that most of the central nodes belong particularly to the DMN. The same for the vulnerability and the strength figures where the prefrontal cortex and the cingulate region are also involved. However, one can notice that the regions that have the highest clustering coefficient do not belong to the DMN.

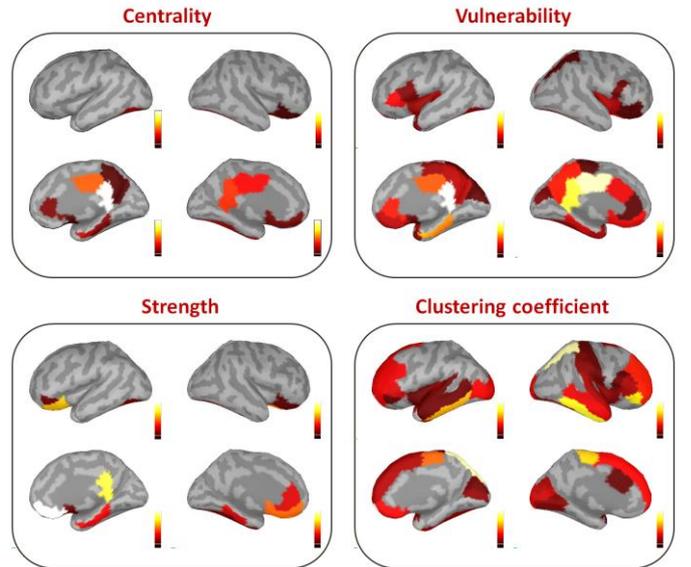

**Figure 2. Graph measures distribution on the cortical surface for the 14 subjects.**

### IV. DISCUSSION

A node has been previously defined as hub if it has an unusually high strength or centrality, and a low clustering coefficient [29]. Based on this definition of "hubness", we can say that the "isthmus cingulate" region plays the role of a hub. Similar findings on the important role of cingulate gyrus region have been reported by many structural and functional analysis based MRI and MEG [16][17][18][19][20]. Examining our obtained results, we also recognize the importance of the orbito-frontal region, already considered as critical [15] [17][18][19]. Moreover, [20] have insisted on the importance of the posterior cingulate also depicted here as central and vulnerable node with low clustering coefficient. In addition, results suggest that the DMN plays a major role in maintaining an efficient brain communication during rest, since the majority of the relevant nodes are included in the DMN. This confirms the fact that the DMN is highly activated during rest [30][8]. Furthermore, EEG source connectivity approach could offer a unique insight into the way the brain network can be dynamically reconfigured and reorganized, thanks to the excellent time resolution offered by EEG. Further work will be the tracking of the dynamic characteristics of RSNs and the analysis of how the dynamic interactions across RSNs are spatially and temporally modified.

## V. CONCLUSION

In this paper, we used the dense EEG source connectivity method to explore the human brain network architecture during rest. The networks were characterized in terms of node's centrality, vulnerability, strength and clustering. Our results confirmed the existence of regions playing the role of "functional hubs", consistent with the state-of-the art findings. Moreover, we reported the critical role of nodes that correspond to the default mode network (DMN). Our findings highlighted also the capacity of EEG source connectivity method to reveal the brain network topology during rest.

ACKNOWLEDGMENT

This work was supported by the Rennes University Hospital (COREC Project named BrainGraph, 2015-17). The work has also received a French government support granted to the CominLabs excellence laboratory and managed by the National Research Agency in the "Investing for the Future" program under reference ANR-10-LABX-07-01. It was also financed by Azm center for research in biotechnology and its applications. Authors would like to thank Dufor O. for his help in the data acquisition.